\documentstyle[12pt]{article}

\def\pagesetup{
\textwidth 6.0in
\textheight 8.5in
\pagestyle{empty}
\topmargin -0.25truein
\oddsidemargin 0.30truein
\evensidemargin 0.30truein
\raggedbottom\parindent=20pt
\baselineskip=14pt}

\pagesetup

\newenvironment{Titlepage}{\parsep=0pt \topsep=0pt}{\relax}

\def\Title#1\endTitle{\begin{center}%
   \baselineskip=16pt 
 \bf #1\\[.5cm]}                    

\def\Author#1#2\endAuthor{\small\it
   {\rm #1}\\[1pt] #2\\[.3cm]}

\def\endAuthors{\end{center}
                \vglue .5cm    
                    }

\newenvironment{Abstract}%
       {\centering\bgroup
          \begin{minipage}{30pc}\small
            \noindent
	    \centerline{ABSTRACT}
	    \parindent=0pt}%
         {\end{minipage}\egroup\par
         \normalsize}

\begin{document}

\begin{Titlepage}
\vskip -0.5truein
\vbox{
\rightline{\small {\bf QMW-PH-94-35}}
\rightline{\small {\bf hepth/9410069}}
}

\bigskip
\Title
DUALITY VERSUS SUPERSYMMETRY\footnote{Contribution to the
Seventh Marcel Grossmann Meeting on General Relativity,
Stanford University,
July 24th---30th, 1994.}
\endTitle

\Author{TOM\'AS ORT\'IN\footnote{E-mail address: {\tt
ortin@qmchep.cern.ch}}}
Department of Physics, Queen Mary and Westfield College,
Mile End Road, London E1 4NS, U.K.
\endAuthor
\endAuthors

\begin{Abstract}

We study the effect of S-duality and target-space duality tranformations
of $N=4,d=4$ and $N=1,d=10$ supersymmetric configurations on their
Killing spinors.  We find that, under reasonable assumptions, the dual
configurations are also supersymmetric and that the Killing spinors
transform in a simple way.

\end{Abstract}

\end{Titlepage}

\vspace{.5cm}

There is an increasing interest in finding new classical backgrounds of
string theory describing black holes, gravitational waves etc.  Many
have been found recently as solutions of the low-energy effective action
and exact CFTs, and some of the most interesting ones have unbroken
supersymmetries.  On the other string theory has several {\it duality}
symmetries that relate very different-looking backgrounds.
T-duality \cite{B} relates two backgrounds with an isometry which are
described by equivalent CFTs.  From a different point of view T-duality
also relates two solutions of the leading order in $\alpha^{\prime}$
effective action with an isometry.  S-duality is essentially a symmetry
of $N=4,d=4$ supergravity \cite{CSF} (the leading order in
$\alpha^{\prime}$ of the effective action in $d=4$) formulated with a
pseudoscalar axion, although it has been conjectured to be an exact
non-perturbative symmetry of string theory \cite{S}.  Our aim is to study
if these duality transformations preserve unbroken supersymmetries and,
if they do, what is the transformation law of the Killing spinors.  We
will do it in the framework of the leading order effective action.

In the case of S-duality our results can be stated as follows \cite{O}:
If $\{g_{\mu\nu},A^{I}_{\mu},\lambda=a+ie^{-2\phi}\}$ is a solution of
$N=4,d=4$ supergravity (Einstein frame) admitting the Killing spinor set
$\epsilon^{I}$, and $\left( \begin{array}{cc}\alpha & \beta \\ \gamma &
\delta \\ \end{array} \right)$ is an $SL(2,Z)$-duality transformation,
then the transformed configuration $\{g_{\mu\nu},A^{\prime
I}_{\mu},\lambda^{\prime}\}$ is also a solution admitting the Killing
spinor
\begin{equation}
\epsilon^{\prime I}=e^{\frac{i}{2}Arg(S)}\epsilon^{I}\, ,
\hspace{1cm}
S=\gamma\lambda+\delta\, .
\end{equation}
To prove this result there is no need to use the equations of motion and
therefore it applies to more general configurations which are not
solutions of the equations of motion. However, if Maxwell's equation is
not satisfied by the configuration then, in general, the Bianchi
identity will not be satisfied after the transformation and one
can not properly speak of an $N=4,d=4$ supergravity configuration.
The S-convariance of the Killing spinor set implies the invariance of
the Bogomolnyi bound in axion-dilaton black holes \cite{O}.

In the case of T-duality our results can be stated as
follows \cite{BKO3}: Let $\{\hat{g}_{\hat{\mu}\hat{\nu}},
\hat{B}_{\hat{\mu}\hat{\nu}}, \hat{\phi}\}$ be a solution of $N=1,d=10$
supergravity admitting the Killing spinor $\epsilon$. If the fields do
not depend on the coordinate $x$ it is natural to use a
Kaluza-Klein-type basis of zehnbeins
$\hat{e}_{\hat{\mu}}{}^{\hat{a}} =\left(
\begin{array}{cc} e_{\mu}{}^{a} & k A_{\mu} \\ 0 & k \\ \end{array}
\right)$. If in this basis $\epsilon$  does not depend on $x$, then
the T-dual configuration with respect to $x$ admits a Killing spinor
$\tilde{\epsilon}$ where
\begin{enumerate}
\item if $x$ is space-like $\tilde{\epsilon}=\epsilon$.
\item if $t$ is time-like $\tilde{\epsilon}=\gamma_{x}\epsilon$.
\end{enumerate}
An interesting example that illustrates this result is provided by the
classes of solutions known as Supersymmetric String Waves and
Generalized Fundamental Strings  \cite{BEKO}.  These two classes of
solutions are related by T-duality and admit Killing spinors.  In a
KK-type basis of zehnbeins the Killing spinors are equal.  In particular
some SSW are dual to the four-dimensional extreme electric dilaton black
hole uplifted to $d=10$ in a way in which the preservation of the
four-dimensional unbroken supersymmetry is guaranteed \cite{BKO2}.

Another example is provided by the four-dimensional extreme magnetic
dilaton black hole uplifted to $d=10$ preserving supersymmetry \cite{KO}.
In $d=10$ its Killing spinors are constant spinors constrained by
$(1\pm\gamma_{1234})\epsilon_{\pm}=0$.  The configuration is invariant
under T-duality in the time direction, and the constraint on the killing
spinors is invariant under multiplication by $\gamma_{0}$, confirming
the above result.

These results can be extended in a number of interesting ways: inclusion
of (non-Abelian) vector fields, continuous $SO(1,2)$ duality etc and
the results will be published elsewere \cite{BKO3}.

\section*{Acknowledgements}

The author is indebted to E. Bergshoeff and R. Kallosh for their
collaboration and to the Physics Departments of University of Groningen
and Stanford University for their hospitality and financial support.

\end{document}